\documentclass[12pt]{article}
\begin{document}
\def\1{\'{\i}}

\centerline{\Large\bf One-parameter Isospectral Special Functions}
\vspace*{1cm}

\centerline{M.A.Reyes$^\dagger$, D.Jimenez$^\dagger$ and H.C.Rosu$^\ddagger$}
\vspace*{2mm}

\centerline{\small\it $^\dagger$Instituto de F\1sica, Universidad de Guanajuato,
Apdo. Postal E143, 37150 Le\'on,Gto., M\'exico}

\centerline{\small\it $^\ddagger$Applied Mathematics, IPICyT, Apdo. Postal 3-74
Tangamanga, San Luis Potos\1, M\'exico}
\vspace*{5mm}

\begin{center}
\small
\begin{minipage}{14cm}

{\bf Abstract}. Using a combination of the ladder operators of Pi\~na [Rev. Mex. F\'{\i}s. {\bf 41} (1995) 913] and the parametric operators of Mielnik [J. Math. Phys. {\bf 25} (1984) 3387]
we introduce second order linear differential equations whose eigenfunctions are
isospectral to the special functions of the mathematical physics and illustrate the method with several key examples.

\bigskip

{\bf Resumen}. Usando una combinaci\'on de los operadores de escalera de Pi\~na [Rev. Mex. F\'{\i}s. {\bf 41} (1995) 913] y de los operadores parametricos de 
Mielnik [J. Math. Phys. {\bf 25} (1984) 3387] introducimos operadores lineales de segundo orden con eigenfunciones que son formas isoespectrales de las 
funciones especiales de la f\1sica matem\'atica y presentamos algunos ejemplos b\'asicos.   

\end{minipage}
\end{center}

\normalsize

\section{Introduction}

The use of the factorization method \cite{factor} proved to be a powerful tool for extending the class of exactly solvable 
Sturm-Liouville problems especially in quantum mechanics, where in the form of {\em supersymmetric quantum mechanics} led to  new potentials, which are
isospectral to a given problem \cite{coop}.
In a
paper by Mielnik \cite{bogd}, the usual factorization operators 
\begin{equation}
a= \frac{1}{\sqrt{2}} \left( \frac{d~}{dx}+ x \right)
\qquad
a^*=\frac{1}{\sqrt2} \left ( -\frac{d~}{dx}+ x \right)
\end{equation}
of the one dimensional harmonic oscillator Hamiltonian
\begin{equation}\label{hosc}
H+\frac{1}{2}=-\frac 12\frac{d^2~}{dx^2}+\frac 12x^2+\frac{1}{2}
\end{equation}
have been replaced by new operators
\begin{equation}
b=\frac{1}{\sqrt{2}} \left ( \frac{d~}{dx} + \beta(x) \right ) \qquad
b^*=\frac{1}{\sqrt2} \left ( -\frac{d~}{dx} + \beta(x) \right )~.
\end{equation}
In order that these new operators factorize 
the same Hamiltonian (\ref{hosc}) 
\begin{equation}\label{fbb*}
bb^*=aa^*=H+\frac{1}{2}
\end{equation}
the function $\beta(x)$ should satisfy a Riccati equation of the form
\begin{equation}
\beta'+\beta^2=1+x^2~.
\end{equation}
Using the solutions of this equation
\begin{equation}\label{beta}
\beta(x)=x+\frac{e^{-x^2}}{\gamma + \int\limits_{0}^x e^{-y^2}dy}
\end{equation}
one can introduce a new Hamiltonian $\widetilde H$, which is defined by the inverse
factorization of (\ref{fbb*})
\begin{equation}\label{fb*b}
b^*b=\widetilde H-\frac{1}{2}=H+\frac{1}{2}-1- \frac{d~}{dx}
\left[\frac{e^{-x^2}}{\gamma +\int\limits_{0}^x e^{-y^2}dy} \right]
\end{equation}
with new potential functions
\begin{equation}
\widetilde V(x)= \frac{x^{2}}{2} - \frac{d~}{dx} \left[
\frac{e^{-x^2}}{\gamma +\int\limits_{0}^x e^{-y^2}dy} \right]
\end{equation}
and whose eigenfunctions
\begin{equation}
\widetilde \psi_{n}=b^*\psi_{n-1}
\qquad (n=1,2,...)
\end{equation}
are isospectral to the harmonic oscillator eigenfunctions $\psi_n$.


\section{Factorization of special functions}

The great majority of differential equations appearing in Physics
can be factorized by means of ladder operators. Therefore, we should be
able to apply the procedure described in the previous section to the raising
and lowering operators of the important class of Sturm-Liouville problems and get in this way
isospectral second order differential equations.  To attain this objective we proceed as
follows.

One of the possible forms of factorizing a subclass of second order differential operators
associated to the special functions of mathematical physics was introduced by
Pi\~na \cite{pina}. Consider the Sturm-Liouville problem
\begin{equation}\label{liou}
{\cal L}_n \psi_n(x)\equiv \left[ P(x) \frac{d^2~}{dx^2} + Q(x) \frac{d~}{dx} +
R_{n}(x) \right] \psi_n(x)=0~,
\end{equation}
where $P$, $Q$ and $R_{n}$ are functions of the variable $x$, and $R_{n}$ depend
on the index $n$.  Then, it is possible to construct raising and lowering
operators \cite{pina}
\begin{equation}\label{a+a-}
A_{n}^+= \sqrt{P} \frac{d~}{dx} + a_{n}^+ \qquad \qquad A_{n}^-= \sqrt{P}
\frac{d~}{dx} + a_{n}^-
\end{equation}
that can factorize  Eq.(\ref{liou}) in two ways
\begin{eqnarray}
A_{n+1}^-A_n^+&=&{\cal L}_n + K_n \label{fac1}\\
A_n^+A_{n+1}^-&=&{\cal L}_{n+1} + K_n \label{fac2}~,
\end{eqnarray}
where the constant $K_n$ is the same in both factorizations.
Furthermore, the functions $a_{n}^+$, $a_{n+1}^-$, turn out to be
\begin{equation}
a_{n+1}^- = \frac{1}{2} \left [ \frac{Q}{\sqrt{P}}- \frac{d~}{dx} \sqrt{P}+
c_{n}+ \int \frac{1}{\sqrt{P}}(R_{n+1} - R_{n}) dx \right ]
\end{equation}
\begin{equation}
a_{n}^+= \frac{1}{2}\left [ \frac{Q}{\sqrt{P}}- \frac{d~}{dx} \sqrt{P}-
c_{n}- \int \frac{1}{\sqrt{P}}(R_{n+1} - R_{n}) dx \right ]~,
\end{equation}
where $c_n$ is an integration constant.  From Eq.(\ref{fac1}), one
may consider the constant $K_n$ as the eigenvalue corresponding to the
eigenfunction $\psi_n$ for the operator $A_{n+1}^-A_n^+$.


Let us now define new operators $B_n^+$, $B_{n+1}^-$ by
\begin{eqnarray}
B_{n}^{+}=A_{n}^{+}+b_{n}^{+}\nonumber \\
B_{n}^{-}=A_{n}^{-}+b_{n}^{-}
\end{eqnarray}
and demand that they can also factorize the Sturm-Liouville operator
${\cal L}_n(x)$ as
\begin{equation}\label{facb}
B_{n+1}^{-}B_{n}^{+}=A_{n+1}^-A_n^+={\cal L}_n(x)+K_n~.
\end{equation}
Then, the following relationship should be fulfilled
\begin{equation}
\sqrt{P}\left(b_{n}^+\frac{d~}{dx}+b_{n+1}^- \frac{d~}{dx}\right)+
\sqrt{P}\frac{d b_n^+}{dx}+
b_{n}^+ a_{n+1}^- +
a_{n}^+ b_{n+1}^- +
b_{n}^+ b_{n+1}^- =0~.
\end{equation}
Therefore, the functions $b_n^+$, $b_{n+1}^-$ must satisfy
\begin{equation}
b_{n+1}^-=-b_{n}^+
\end{equation}
\begin{equation}\label{ricb}
\sqrt{P}\frac{db_n^+}{dx}-{b_{n}^{+}}^{2}+b_{n}^{+}(a_{n+1}^{-}-a_{n}^{+})=0~.
\end{equation}
Eq.(\ref{ricb}) is a Riccati type equation,
which can be easily solved to get
\begin{equation}\label{defb}
b_{n}^+(x)=\frac{e^{\delta(x)}}{\gamma-\int\limits_{x_0}^x
\frac{e^{\delta(y)}}{\sqrt{P(y)}}dy}~.
\end{equation}
Here, $\delta(x)$ is defined by the indefinite integral
\begin{equation}\label{alfa}
\delta(x) \equiv \int\limits^x \frac{(a_{n}^+(y)-
a_{n+1}^-(y))}{\sqrt{P(y)}} dy
\end{equation}
and $\gamma$ is an integration constant. $x_0$ may be chosen as the
point where the integrand vanishes.


Similarly to Mielnik's new Hamiltonian $\widetilde H$, we now
introduce the second order differential operator $\widetilde{\cal L}_n(x)$
given by
\begin{equation}
\widetilde{{\cal L}}_{n+1}= B_n^+B_{n+1}^--K_n = {\cal L}_{n+1}
-2\sqrt{P}\frac{db_{n}^+}{dx}~.
\end{equation}
That is, the new operator $\widetilde{\cal L}_n$ differs from ${\cal L}_n$ by
the derivative of the solution of the Riccati equation (\ref{ricb}), in the same way as
Mielnik's Hamiltonian $\widetilde H$ differs from the harmonic oscillator
Hamiltonian $H$, as can be seen in Eq.(\ref{fb*b}).

If we now define the functions
\begin{equation}
\widetilde{\psi}_{n+1} \equiv B_n^+\psi_{n}
\qquad n=0,1,2,\ldots
\end{equation}
where $\psi_{n}$ are the eigenfunctions of ${\cal L}_n$, we can see that
\begin{equation}
\widetilde{\cal L}_{n+1}\widetilde{\psi}_{n+1}=
\left(B_{n}^+B_{n+1}^--K_n\right) B_{n}^+\psi_{n}=
B_{n}^+{\cal L}_n\psi_n=0
\end{equation}
Therefore, these $\widetilde{\psi}_n$ are the eigenfunctions for the new
operator $\widetilde{\cal L}_n$, and we can write the eigenvalue equation
\begin{equation}
B_n^+B_{n+1}^-\widetilde\psi_{n+1}=
\left( \widetilde{\cal L}_{n+1}+K_n \right) \psi _{n+1}=
K_{n}\widetilde\psi _{n+1}
\end{equation}

As can be seen from Eq.(\ref{defb}), we have constructed a one paramenter
family of operators $\widetilde{\cal L}_n(x;\gamma)$ which are ``isospectral"
to the original Sturm-Liouville operator ${\cal L}_n(x)$, for the allowed
values of the parameter $\gamma$ that produce a non-divergent function
$b_n^+(x)$.

As we can see, $B_n^+,\ B_{n+1}^-$ are not ladder operators as are $A_n^+,\
A_{n+1}^-$.  However, one can easily verify that the third order operators
\begin{eqnarray}
C_n^+ &\equiv& B_n^+A_{n-1}^+B_n^- \\
C_{n+1}^- &\equiv& B_{n-1}^+A_n^-B_{n+1}^-
\end{eqnarray}
play the role of raising and lowering operators, respectively, for the
functions $\widetilde\psi_n(x)$.  The use of ladder operators of order
higher than two is not easy to find in the literature, but some work in this
direction has already been reported \cite{abdl}.

Since $B_{n+1}^-$, and therefore $C_{n+1}^-$, is not defined for $n\!=\!0$, the
space defined by the eigenfunctions $\widetilde\psi_n$ lack the element with
$n\!=\!0$, similarly to what happens in SUSY factorization \cite{coop}.


\section{Examples}

Here, we proceed to find the new operators $\widetilde{\cal L}_n(x;\gamma)$ and
their eigenfunctions, from the factorizations of the special functions of
mathematical physics.


\subsection{Hermite polynomials}

The Hermite differential equation
\begin{equation}
\frac{d^2H_{n}(x)} {dx^2} - 2x\frac{dH_n(x)}{dx} +2nH_n(x) =0
\end{equation}
has raising and lowering differential operators given by
\begin{equation}
\left( \frac{d~}{dx}-2x \right) H_n(x)=-H_{n+1}
\end{equation}
\begin{equation}
\frac{d~}{dx} H_{n+1}(x)=2(n+1)H_{n}(x)~.
\end{equation}
In this case, the $\delta$-integral in Eq.(\ref{alfa}) is
\[ \delta =\int\limits^x (-2y)dy= -x^{2} \]
and therefore
\begin{equation}
b_{n}^{+}=\frac{e^{-x^{2}}}{\gamma -\int\limits_0^x e^{-y^{2}}dy}~.
\end{equation}
The integrand in the denominator being positive definite, one should impose the condition 
$\left| \gamma \right|  >\frac{\sqrt{\pi }}{2}$ \cite{bogd} in order to have a well-defined operator.

Now, with the use of Eq.(\ref{ricb}) one gets 
\begin{equation}
\frac{db_{n}^+}{dx} =
\left( \frac{e^{-x^{2}}}{\gamma -\int\limits^x_0 e^{-y^{2}}dy}\right) ^{2}-
\frac{2xe^{-x^{2}}}{\gamma -\int\limits^x_0 e^{-y^{2}}dy}~,
\end{equation}
and therefore, the second order differential operator
\begin{equation}
\widetilde{\cal L}_{n+1}(x;\gamma)=
\frac{d^2~}{dx^{2}}-2x\frac{d~}{dx}+2n
+\frac{4xe^{-x^{2}}}{\gamma -\int\limits^x_0 e^{-y^{2}}dy}
-\frac{2e^{-2x^{2}}}{\left(\gamma -\int\limits^x_0 e^{-y^{2}}dy\right)^{2}}
\end{equation}
has parametric eigenfunctions given by
\begin{equation}
{\widetilde H}_{n+1}(x;\gamma)=-H_{n+1}(x)
+\frac{e^{-x^{2}}}{\gamma-\int\limits_{0}^{x}e^{-y^{2}}dy} H_{n}(x)~.
\end{equation}


\subsection{Laguerre polynomials}

For the Laguerre differential equation
\begin{equation}
x^2 \frac{d^2L_{n}^\alpha(x)}{dx^2} + [(\alpha + 1)x-x^2]
\frac{dL_{n}^\alpha(x)}{dx}+
nxL_{n}^\alpha(x)=0~,
\end{equation}
the raising and lowering operators are
\begin{equation}
\left( x \frac{d~}{dx} + \alpha + n+ 1 - x \right) L_{n}^\alpha(x)=
(n+1)L_{n+1}^\alpha(x)
\end{equation}
\begin{equation}
\left( x \frac{d~}{dx}-n-1 \right)L_{n+1}^\alpha(x)=
-(\alpha +n +1)L_{n}^\alpha(x)~.
\end{equation}
The $\delta$-integral is
\begin{equation}
\delta=\int\limits^x \frac{\alpha +2(n+1)-y}{y}dy
=\ln x^{[\alpha +2(n+1)]}-x
\end{equation}
and hence
\begin{equation}
b_{n}^{+}=\frac{x^{\alpha+2n+2}e^{-x}}{\gamma -\int\limits_0^x
y^{(\alpha +2n+1)} e^{-y}dy}~.
\end{equation}
For $x>0$, the integral in the denominator is positive definite, with maximum
value
\[
\int\limits_{0}^{\infty }y^{(\alpha +2n+1)} e^{-y}dy=\Gamma(\alpha+2n+2)
\]
and, since there is no upper limit for increasing $n$, we must have $\gamma<0$.

The second order differential operator
\begin{eqnarray}
\widetilde {\cal L}_{n+1}(x;\gamma)&=&
x^2\frac{d^2~}{dx^2}+\left[ (\alpha+1)x-x^2\right] \frac{d~}{dx}+nx+
\frac{2[x-(\alpha+2n+2)]x^{\alpha+2n+1}e^{-x}}
{\gamma-\int\limits_0^x y^{\alpha+2n+1}e^{-y}dy} \nonumber \\
&-&
\frac{2x^{2\alpha+4n+2}e^{-2x}}
{\left( \gamma-\int\limits_0^x y^{\alpha+2n+1}e^{-y}dy \right)^2}
\end{eqnarray}
has as parametric eigenfunctions
\begin{equation}
\widetilde L_{n+1}^\alpha(x;\gamma)=
(n+1)L_{n+1}^\alpha(x)+
\frac{x^{\alpha+2n+2}e^{-x}}{\gamma-\int\limits_0^x y^{\alpha+2n+1}e^{-y}dy}
L_n^\alpha~.
\end{equation}


\subsection{Legendre polynomials}

The Legendre differential equation
\begin{equation}
(x^2-1)^2 \frac{d^2P_{n}(x)}{dx^2}+2x(x^2-1)\frac{dP_{n}(x)}{dx}%
-n(n+1)(x^2-1)P_{n}(x) =0
\end{equation}
has the raising and lowering operators
\begin{equation}
\left[ (x^2-1) \frac{d~}{dx}+(n+1)x \right] P_{n}(x)=(n+1)P_{n+1}(x)
\end{equation}
\begin{equation}
\left[ (x^2-1)\frac{d~}{dx}-(n+1)x \right]P_{n+1}(x)= -(n+1)P_{n}(x)~.
\end{equation}
The $\delta$-integral is in this case 
\begin{equation}
\delta =-2(n+1)\int\limits^x \frac{y}{1-y^{2}}dy=\ln (1-x^{2})^{n+1}
\end{equation}
hence
\begin{equation}\label{bleg}
b_{n}^{+}=\frac{(1-x^{2})^{n+1}}{\gamma -\int\limits_{-1}^x (1-y^{2})^{n}dy}~.
\end{equation}
The integral in the denominator is positive definite, which, for a given $n$,
has maximum value
\[
\int\limits_{-1}^{+1} (1-x^{2})^{n}dx=
2\int\limits_0^{\pi/2}\mbox{sen}^{2n+1}\theta\, d\theta=
\frac{2(2n)!!}{(2n+1)!!}
\]
and, therefore, $|\gamma|>2$.

Hence, the second order differential operator
\begin{eqnarray}
\widetilde {\cal L}_{n+1}(x;\gamma)=
(1-x^2)\frac{d^2~}{dx^2}-2x\left(1-x^2\right)\frac{d~}{dx}+n(n+1)(1-x^2)+
\nonumber \\
\frac{4(n+1)(1-x^2)^{n+1}}{\gamma-\int\limits_{-1}^x (1-y^2)^ndy}-
\frac{2(1-x^2)^{2n+2}}{\left(\gamma-\int\limits_{-1}^x (1-y^2)^ndy\right)^2}
\end{eqnarray}
has as parametric eigenfunctions
\begin{equation}
\widetilde P_{n+1}(x;\gamma)=
-(n+1)P_{n+1}(x)+\frac{(1-x^2)^{n+1}}{\gamma-\int\limits_{-1}^x (1-y^2)^ndy}
P_n(x)~.
\end{equation}


\subsection{Chebyshev polynomials}

For the Chebyshev differential equation
\begin{equation}
(1-x^2)^2 \frac{d^2T_{n}(x)}{dx^2}-x(1-x^2)\frac{dT_{n}(x)}{dx}
+n^2(1-x^2)T_{n}(x)=0~,
\end{equation}
the corresponding raising and lowering operators are
\begin{equation}
\left[ (1-x^2)\frac{d~}{dx} - nx \right] T_{n}(x)=-nT_{n+1}(x)~.
\end{equation}
\begin{equation}
\left[ (1-x^2)\frac{d~}{dx}+(n+1)x \right] T_{n+1}(x)=(n+1)T_{n}(x)~.
\end{equation}
In this case, the $\delta$-integral is
\begin{equation}
\delta =-(2n+1)\int\limits^x \frac{y}{1-y^2}dy=
\left(n+\frac{1}{2}\right) \ln (1-x^2)
\end{equation}
and, hence
\begin{equation}
b_{n}^{+}=\frac{(1-x^{2})^{n+\frac{1}{2}}}
{\gamma -\int\limits_1^x (1-y^2)^{n-\frac{1}{2}}dy}~.
\end{equation}
The integral in the denominator has maximum value
\[
\int\limits_{-1}^1 (1-x^2)^{n-\frac{1}{2}}dx=
2\int\limits_0^{\pi/2}\mbox{sen}^{2n}\theta d\theta=
\frac{\pi(2n-1)!!}{(2n)!!}~,
\]
which imposes the condition $\gamma>\pi$.

We can thus construct the second order differential operator
\begin{eqnarray}
\widetilde {\cal L}_{n+1}(x;\gamma)=
(1-x^2)\frac{d^2~}{dx^2}-x\left(1-x^2\right)\frac{d~}{dx}+n^2(1-x^2)+
\nonumber \\
\frac{2(2n+1)x(1-x^2)^{n+\frac{1}{2}}}
{\gamma-\int\limits_{-1}^x (1-y^2)^{n-\frac{1}{2}}dy}-
\frac{2(1-x^2)^{2n+1}}
{\left(\gamma-\int\limits_{-1}^x (1-y^2)^{n-\frac{1}{2}}dy\right)^2}
\end{eqnarray}
with the parametric eigenfunctions
\begin{equation}
\widetilde T_{n+1}(x;\gamma)=-nT_{n+1}(x)+
\frac{(1-x^2)^{n+\frac{1}{2}}}
{\gamma-\int\limits_{-1}^x (1-y^2)^{n-\frac{1}{2}}dy} T_n(x)~.
\end{equation}


\subsection{Jacobi functions}

The equation defining the Jacobi functions
\begin{equation}
x^2 (1-x)^2 \frac{d^2f_{n}(x)}{dx}+x(1-x)[\lambda-(\alpha+1)x]\frac{df_{n}(x)%
}{dx} + x(1-x)n(n+\alpha)f_{n}(x)=0
\end{equation}
has the raising and lowering operators
\begin{equation}
\left[x(1-x)\frac{d~}{dx}-(n+\alpha)\left(x-\frac{n+\lambda} {2n+\alpha+1}%
\right) \right]f_{n}(x)=\frac{(n+\alpha)(n+\lambda)} {2n+\alpha+1}f_{n+1}(x)
\end{equation}
\begin{equation}
\left[ x(1-x)\frac{d~}{dx} + (n+1) \left( x- \frac{n+1+\alpha-\lambda}{%
2n+\alpha+1} \right) \right]f_{n+1}(x)=
-\frac{(n+1)(n+1+\alpha-\lambda)}{2n+\alpha+1}f_{n}(x)~.
\end{equation}
Hence, the $\delta$-integral becomes
\[
\delta =\int\limits^x \frac{u-vy}{y(1-y)}dy=u\ln x+(v-u)\ln (1-x)~,
\]
where
\[
u=\frac{(n+\alpha)(n+\lambda)+(n+1)(n+1+\alpha-\lambda)}
{2n+\alpha +1} \ , \qquad
v=2n+\alpha +1~.
\]
Therefore, we find that
\begin{equation}
b_{n}^{+}=\frac{x^{u}(1-x)^{v-u}}{\gamma-
\int\limits^x y^{u-1}(1-y)^{v-u-1}dy}~,
\end{equation}
where the integral in the denominator can be written in terms of the incomplete Beta
functions $B_{-1}(u,v-u)$ and $B_x(u,v-u)$.  In the particular case $v>u$, to
obtain the range of allowed values for the $\gamma$ parameter, we see that
\begin{eqnarray}
\int\limits_{-1}^1 x^{u-1}(1-x)^{v-u-1}dx=
-\int\limits_0^{-1} x^{u-1}(1-x)^{v-u-1}dx+
\int\limits_0^1 x^{u-1}(1-x)^{v-u-1}dx< \nonumber \\
< 2\int\limits_0^1 x^{u-1}(1-x)^{v-u-1}dx=
2B(u,v-u)=\frac{2\Gamma(v-u)\Gamma(u)}{\Gamma(v)}~.
\end{eqnarray}
Therefore, we have to demand 
$|\gamma|>\frac{2\Gamma(v-u)\Gamma(u)}{\Gamma(v)}$.

The second order differential operator
\begin{eqnarray}
\widetilde {\cal L}_{n+1}(x;\gamma)&=&
x^2(1-x^2)\frac{d^2~}{dx^2}+
x(1-x)\left(\lambda-(\alpha+1)x\right)\frac{d~}{dx}+
x(1-x)n(n+\alpha)+
\nonumber \\
&+&\frac{2(vx-u)x^u(1-x)^{v-u}}
{\gamma-\int\limits_{-1}^x y^{u-1}(1-y)^{v-u-1}dy}-
\frac{2x^{2u}(1-x)^{2(v-u)}}
{\left(\gamma-\int\limits_{-1}^x y^{u-1}(1-y)^{v-u-1}dy\right)^2}
\end{eqnarray}
has the parametric eigenfunctions
\begin{equation}
\widetilde f_{n+1}(x;\gamma)=
\frac{(n+\alpha)(n+\lambda)}{2n+\alpha+1}f_{n+1}(x)+
\frac{x^u(1-x)^{v-u}}
{\gamma-\int\limits_{-1}^x y^{u-1}(1-y)^{v-u-1}dy} f_n(x)~.
\end{equation}


\subsection{Jacobi polynomials}

The differential equation defining the Jacobi polynomials
\begin{eqnarray}
(1-x^2)^2\ \frac{d^2P_{n}^{\alpha\beta}(x)}{dx^2}&+&
(1-x^2)[\beta-\alpha-(\alpha+\beta+2)x]
\frac{dP_{n}^{\alpha\beta}(x)}{dx}+
\nonumber \\ & &
+(1-x^2)n(n+\alpha+\beta+1)P_{n}^{\alpha\beta}(x)=0
\end{eqnarray}
has raising and lowering operators given by
\begin{eqnarray}
\left[(1-x^2)\frac{d~}{dx}\right.&+&
\left. (n+1+\alpha+\beta) \left( -x+ \frac{\beta-\alpha}
{2n+2+\alpha+\beta} \right) \right] P_{n}^{\alpha\beta}(x) =
\nonumber \\ & &
= -\frac{2(n+1)(n+1+\alpha+ \beta)}{2n+2+\alpha+\beta}P_{n+1}^{\alpha\beta}(x)
\end{eqnarray}
\begin{eqnarray}
\left[(1-x^2)\frac{d~}{dx}\right. &+&
\left. (n+1) (x+\frac{\beta- \alpha}{2n+2+\alpha+\beta}) %
\right] P_{n+1}^{\alpha\beta}(x) =
\nonumber \\ & &
= \frac{2(n+1+\alpha)(n+1+\beta)}{2n+2+\alpha+\beta} P_{n}^{\alpha\beta}(x)~.
\end{eqnarray}
In this case, we have 
\[
\delta = \int\limits^x \frac{p-qy}{1-y^2}dy=
\frac{1}{2}(q+p)\ln (1+x)+\frac{1}{2}(q-p)\ln(1-x)~,
\]
where
\[
p=\frac{\beta^2-\alpha^2}{2n+2+\alpha+\beta}
\ , \qquad
q=2n+2+\alpha +\beta~.
\]
Hence
\begin{equation}
b_{n}^{+}=\frac{(1+x)^{\frac{1}{2}(q+p)}(1-x)^{\frac{1}{2}(q-p)}}
{\gamma -\int\limits^x_{-1}
(1+y)^{\frac{1}{2}(q+p)-1}(1-y)^{\frac{1}{2}(q-p)-1}dy}~.
\end{equation}
For the parameter $\gamma$, since $q>p$, we demand that
\[
\gamma>
\int\limits_{-1}^1 (1+x)^{\frac{1}{2}(q+p)-1}(1-x)^{\frac{1}{2}(q-p)-1}dx=
2^{q-1}\frac{\Gamma(\frac{q+p}{2})\Gamma(\frac{q-p}{2})}{\Gamma(q)}~.
\]

From here, we construct the second order differential operator
\begin{eqnarray}
\widetilde {\cal L}_{n+1}(x;\gamma)&=&
(1-x^2)^2\frac{d^2~}{dx^2}+
(1-x^2)\left[\beta-\alpha-(\alpha+\beta+2)x\right]\frac{d~}{dx}+ \nonumber \\
&+&(1-x^2)n(n+\alpha+\beta+1)+
\frac{2(qx-p)(1+x)^{\frac{1}{2}(q+p)}(1-x)^{\frac{1}{2}(q-p)}}
{\gamma-\int\limits_{-1}^x
(1+y)^{\frac{1}{2}(q+p)-1}(1-y)^{\frac{1}{2}(q-p)-1}dy}- \nonumber \\
&-& \frac{2(1+x)^{q+p}(1-x)^{q-p}}
{\left(\gamma-\int\limits_{-1}^x
(1+y)^{\frac{1}{2}(q+p)-1}(1-y)^{\frac{1}{2}(q-p)-1}dy\right)^2}~,
\end{eqnarray}
whose parametric eigenfunctions are
\begin{eqnarray}
\widetilde P_{n+1}^{\alpha\beta}(x;\gamma) &=&
-\frac{2(n+1)(n+1+\alpha+ \beta)}{2n+2+\alpha+\beta}P_{n+1}^{\alpha\beta}(x)
\nonumber \\ &+&
\frac{(1+x)^{\frac{1}{2}(q+p)}(1-x)^{\frac{1}{2}(q-p)}}
{\gamma -\int\limits^x_{-1}
(1+y)^{\frac{1}{2}(q+p)-1}(1-y)^{\frac{1}{2}(q-p)-1}dy} P_n^{\alpha\beta}(x)~.
\end{eqnarray}


\subsection{Bessel functions}

For the Bessel differential equation
\begin{equation}
\frac{d^2 J_{n}(x)}{dx^2} + \frac{1}{x} \frac{dJ_{n}(x)}{dx} + ( 1- \frac{n^2%
}{x^2}) J_{n}(x) =0~,
\end{equation}
the raising and lowering operators are
\begin{equation}
\left( \frac{d~}{dx} - \frac{n}{x} \right ) J_{n}(x) = - J_{n+1}(x)
\end{equation}
\begin{equation}
\left( \frac{d~}{dx} + \frac{n+1}{x} \right ) J_{n+1}(x) = J_{n}(x)~.
\end{equation}

The $\delta$-integral is found to be
\[
\delta =\int\limits^x \left(-\frac{n}{y}-\frac{n+1}{y}\right) dy=
\ln x^{-(2n+1)}~,
\]
and therefore
\begin{equation}
b_{n}^{+}=\frac{x^{-(2n+1)}}{\gamma' -\int\limits_\infty^x y^{-(2n+1)}dy}
=\frac{2n}{\gamma x^{2n+1}+x}
\end{equation}
with $\gamma=2n\gamma'\ge 0$.

From here, we can construct a second order differential operator defined by
\begin{equation}
\widetilde{\cal L}_{n+1}(x;\gamma)=
\frac{d^2~}{dx^{2}}+\frac{1}{x}\frac{d~}{dx}+
\left(1-\frac{(n+1)^{2}}{x^{2}}\right)
+\frac{4n+4n\gamma(2n+1) x^{2n-1}}{\left(\gamma x^{2n+1}+x\right)^{2}}~,
\end{equation}
whose parametric eigenfunctions are given by
\begin{equation}
{\widetilde J}_{n+1}(x;\gamma)=-J_{n+1}(x)+\frac{2n}{\gamma x^{2n+1}+x}J_n(x)~.
\end{equation}
Hence, in the case of Bessel functions the new eigenfunctions ${\widetilde
J}_{n+1}(x;\gamma)$ are not regular at $x\!=\!0$, except in the case
$\gamma\!=\!0$, for which $\widetilde{\cal L}_{n+1}(x;\gamma)={\cal L}_{n-1}$
and ${\widetilde J}_{n+1}(x;\gamma)=J_{n-1}(x)$.


\section{Conclusion}

We elaborated here on a combination of a class of Sturm-Liouville ladder operators and one-parameter operators of Mielnik type
that allowed us to construct isospectral Sturm-Liouville second-order linear differential operators with parametric eigenfunctions.
Calculations are worked out for a few important cases.


\end{document}